\documentclass[12pt]{article}
\usepackage{cglw}
\usepackage{psfig}
\usepackage{pstricks}
\def\trace{\mathop{\rm trace}}
\catcode`@=11
\def\fnum@figure{{\bf Figure\space\thefigure.}}
\long\def\@makecaption#1#2{%
  \vskip\abovecaptionskip
  \sbox\@tempboxa{#1 #2}%
  \ifdim \wd\@tempboxa >\hsize
    #1 #2\par
  \else
    \global \@minipagefalse
    \hb@xt@\hsize{\hfil\box\@tempboxa\hfil}%
  \fi
  \vskip\belowcaptionskip}
\catcode`@=12

\begin{document}

\baselineskip=24pt

\begin{center}
{\bf Nonparametric Inference for the Cosmic Microwave Background}\\
\medskip
{\sc Christopher R. Genovese},\footnote{
Research supported by NSF Grants SES 9866147 and NSF-ACI-0121671.}
{\sc Christopher J. Miller},\footnote{
Research supported by NSF-ACI-0121671.}
{\sc Robert C. Nichol},\footnote{
Research supported by NSF Grant DMS-0101360.}
{\sc Mihir Arjunwadkar},\footnote{
Research supported by NSF Grant DMS-0101360.}
{\sc and} {\sc Larry Wasserman}\footnote{
Research supported by NIH Grant R01-CA54852-07
and NSF Grants DMS-98-03433, DMS-0104016, and NSF-ACI-0121671.}\\
\medskip
\emph{Carnegie Mellon University}
\end{center}

\begin{quote}
The Cosmic Microwave Background (CMB), which permeates the entire Universe,
is the radiation left over from just 380,000 years after the Big Bang.
On very large scales, the CMB radiation field is smooth and isotropic,
but the existence of structure in the Universe -- 
stars, galaxies, clusters of galaxies, $\ldots$ -- suggests
that the field should fluctuate on smaller scales.
Recent observations,
from the Cosmic Microwave Background Explorer to the Wilkinson Microwave Anisotropy Project,
have strikingly confirmed this prediction.

CMB fluctuations provide clues
to the Universe's structure and composition shortly after the Big Bang
that are critical for testing cosmological models.
For example, CMB data can be used to determine what portion of the Universe is
composed of ordinary matter versus the mysterious dark matter and dark energy.
To this end, cosmologists usually summarize the fluctuations by the power spectrum,
which gives the variance as a function of angular frequency.
The spectrum's shape, and in particular the location and height of its peaks, 
relates directly to the parameters in the cosmological models.
Thus, a critical statistical question is how accurately can these peaks
be estimated.

We use recently developed techniques to construct a nonparametric confidence set for
the unknown CMB spectrum.
Our estimated spectrum, based on minimal assumptions, closely matches the
model-based estimates used by cosmologists, but we can make a wide range
of additional inferences.
We apply these techniques to test various models and to extract confidence
intervals on cosmological parameters of interest.
Our analysis shows that, even without parametric assumptions, the first peak
is resolved accurately with current data but that the second and third peaks
are not.

\end{quote}


\newpage

\section{Introduction}\label{sec::intro}

The ``Big Bang'' model is misnamed,
as one might expect when a term is coined as an insult.
Cosmologist Fred Hoyle first used the name in a BBC radio interview to denigrate the theory,
which opposed the then-dominant Steady State model.
The name Big Bang stuck, as did its evocation of a mighty explosion in space.
But the image of an \emph{explosion} is highly misleading.
What the model actually posits is that the Universe began hot, dense, and expanding.

Within the first second, roughly 13.7 billion years ago,
the Universe achieved temperatures on the order of 1 trillion Kelvin (K, degrees above absolute zero; Schwarz, 2003).
The density during that second was high enough to stop neutrinos, 
which interact so weakly with matter that
they can pass unmolested through a quadrillion kilometers of lead.
What ties this hot, dense beginning to the Universe we see today is expansion.
A useful metaphor for the expanding universe is the surface of an inflating balloon.
As the balloon inflates, space-time itself is stretched; every point moves away from every other point.
Density falls as the universe expands.
If you picture a wave oscillating over the surface of the balloon, the wavelength increases.
Increasing the wavelength of light corresponds to reducing its temperature.
The Universe thus cools as it expands.

Within the first three minutes, the Universe's temperature was over one billion K.
The energy density in space was so high that atoms could not form.
Space was filled with a stew of photons, baryons (e.g., protons and neutrons),
electrons, neutrinos, and other matter.
As the temperature cooled below 1 billion K, light-element 
nuclei (deuterium, helium, some lithium) formed as well, 
in proportions that fit well with observations.
During this period, photons (radiation) were the dominant form of energy in the Universe.
Any fluctuations in density caused by gravity (which affects light and matter) were 
quickly smoothed out and so
could not grow.

When the temperature of the primordial photons had fallen below approximately 12,000 K, 
photons were no longer dominating the interactions among all particles.
Photons and baryons became coupled in a mathematically perfect fluid, while exotic kinds of matter 
began to clump under the influence of gravity.
The interaction between this photon-baryon fluid and such gravitational overdensities 
are of critical importance and will be described below.

When the temperature reached about 3000 K, roughly 380,000 years after the Big Bang,
electrons and protons could combine to form atoms.
This decoupled the photon-baryon fluid, and the photons flew free through space.
This period is named \emph{recombination} and happened, in cosmic terms, very quickly.
After another 200 million years, hydrogen formed after recombination had clumped enough
for the first stars to form, which began the synthesis of heavy elements and the formation of
galaxies that we see today.

Most of the photons released at recombination have travelled through space 
for billions of years without interacting with matter.
The temperature of these primordial photons 
has now cooled to about 2.7K, barely above absolute zero, which puts them in the microwave 
part of the electromagnetic spectrum.
This primordial radiation field, which still pervades the Universe,
is called the Cosmic Microwave Background (CMB).
The CMB thus provides a snapshot of the moment of recombination,
and fluctuations in the temperature across the sky contain information about
the physics of the early universe.

\subsection{The Cosmic Microwave Background Radiation}

\begin{figure}[t]
\vbox
{
  \psfig{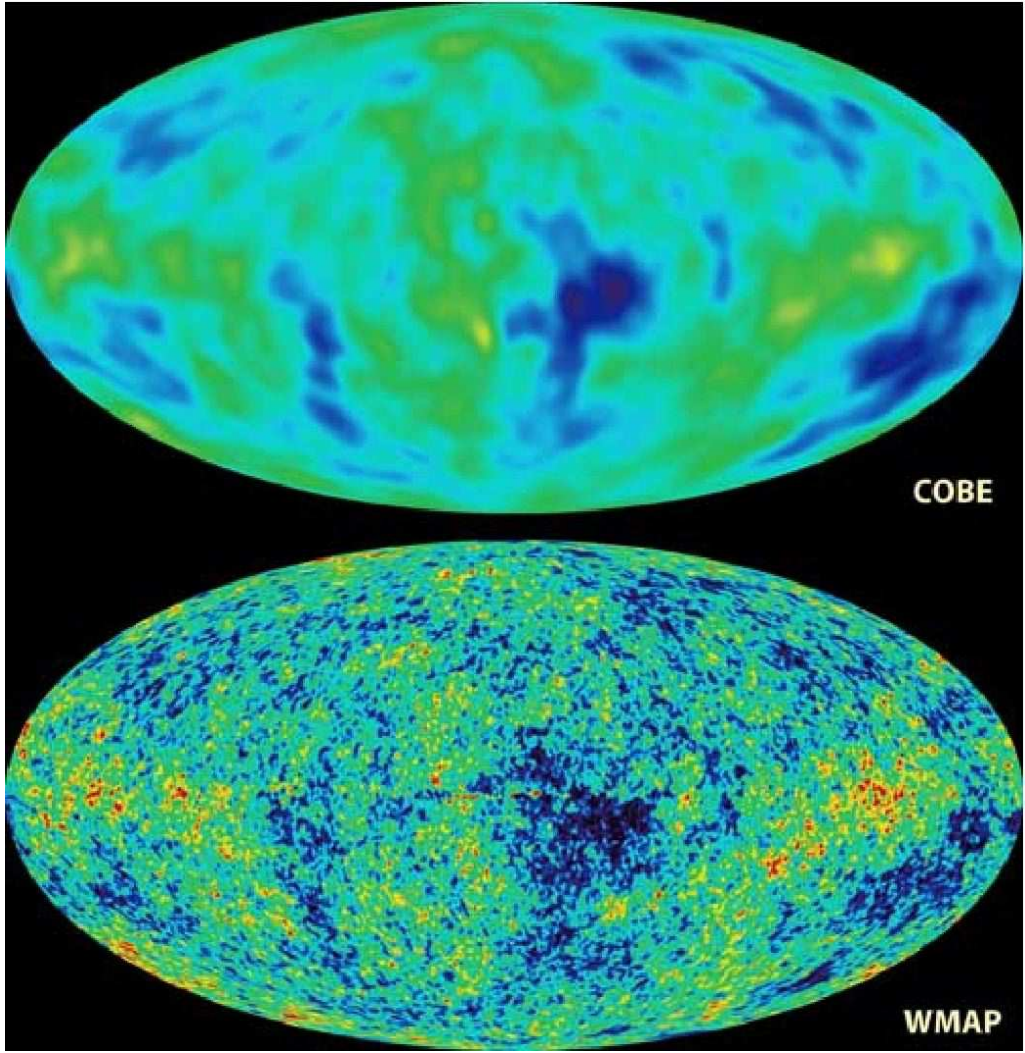}
}
\caption{(top) The CMB as seen by the COBE satellite. The
angular resolution of the satellite is about $10^{\circ}$ and the 
various shades correspond to hot and cool spots with respect to the CMB blackbody
temperature. 
(bottom) 
The CMB from the Wilkinson Microwave Anisotropy Probe. Notice the high angular resolution.
Also notice that the large-scale structures are apparent in both the COBE and the WMAP data.
Image courtesy of the WMAP Science Team and available at the WMAP Mission website:
{\tt http://map.gsfc.nasa.gov}.}
\label{fig::cobeWMAP}
\end{figure}

As we will explain in the remainder of this section,
the temperature fluctuations in the CMB give a snapshot of the physics in the early Universe
and provide critical tests of cosmological models.
In 1992, the Cosmic Microwave Background Explorer (COBE) satellite discovered
fluctuations in the blackbody temperature of the CMB (Smoot et al. 1992). 
These fluctuations are small:
approximately one thousandth of the mean temperature over the sky.
Indeed, almost thirty years of experiments since the CMB's discovery could not detect
any deviation from uniformity.
During the ten years following COBE, 
many more refined measurements were taken; notable experiments include 
MAXIMA, DASI, BOOMERANG (Lee et al. 2002, Halverson et al. 2002, Netterfield et al. 2002).
In 2003, the Wilkinson Microwave Anisotropy Project (WMAP) considerably refined the picture,
increasing spatial resolution by a factor of 33
and sensitivity by a factor of 45 over COBE (Bennett et al. 2003).
In Figure \ref{fig::cobeWMAP}, 
we compare the COBE and WMAP temperature sky maps after removing the mean temperature
$T=2.726$ Kelvin and adjusting for
the motion of our galaxy through the Universe.
The fluctuations' magnitudes are just right to explain the large-scale structure
in the Universe we see today.
For example, if they had been much smaller, there would not be enough local concentration
of mass to seed the formation of galaxies, galaxy clusters, et cetera.

Perhaps the most important summary of the temperature measurements used by cosmologists
is the power spectrum, which gives the temperature variance as a function of spatial frequency.
The spectrum's shape, and in particular the location and height of its peaks, 
relates directly to the parameters in cosmological models.
(See Appendix 2 for a description of these parameters.)
Thus, a critical statistical question is how accurately can these peaks
be estimated.
Of particular interest are the height and location of the first peak
and the relative heights of the successive peaks.

\begin{figure}[t]
\vbox
{
  \psfig{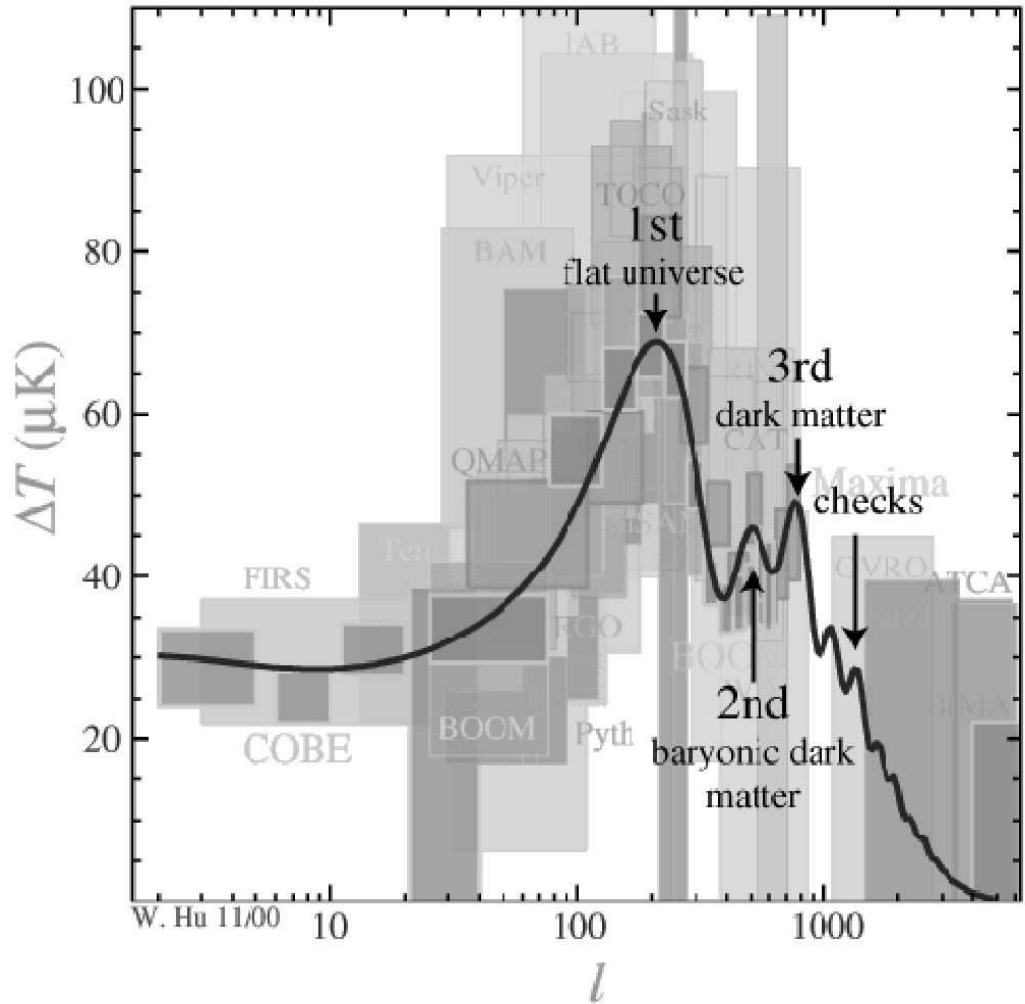}
}
\caption{Estimated CMB spectrum showing the three peaks of interest.
The underlayed boxes give the data ranges and uncertainties 
from a variety of older CMB experiments, not including WMAP.
From Hu (2001).}
\label{fig::cosmospec}
\end{figure}

Figure \ref{fig::cosmospec} displays an estimated spectrum commonly used by cosmologists
and highlights the peaks of interest.
We will give a more precise definition of the spectrum in Section \ref{sec::spectrum},
but here we want to explain how the spectrum's shape relates to the physics
in the time up to recombination.

A key to understanding the physics before recombination is, as mentioned earlier,
that photons and baryons became coupled into a (perfect) fluid.
Mathematical techniques for studying fluid dynamics apply well in this scenario
and have been investigated by many authors (see for instance Hu and Sugiyama 1995; Hu 1999, 2001, 2003; Hu and Dodelson 2002).
The properties of the fluid are determined by the relative density of photons and baryons in the fluid.
Photons provided pressure, and the baryons provided inertia.
As the fluid falls into a gravitational potential well around a clump of higher density,
the pressure from the photons resists compression and the inertia of the baryons increases it.
(Large, isolated potential wells were likely rare in the early universe;
instead, there were random density fluctuations at many scales.)
The result is an oscillation that produces pressure waves -- sound -- in the photon-baryon fluid.
These \emph{acoustic oscillations} account for much of the interesting structure
in the spectrum, particularly the size and arrangement of peaks.
The imprint of those waves remains in the CMB as a pattern of hot and cold spots.

To understand the peaks in the power spectrum, it is helpful to decompose the acoustic
oscillations into their basic components, or modes.
The first peak of the spectrum represents the fundamental tone of the
oscillations, and the other peaks in the spectrum represent harmonics
of this tone.  The fundamental corresponds to the mode for which one 
compression occurs between the Big Bang and recombination.  
Each successive harmonic corresponds to an additional
half-cycle, compressions followed by rarefication (decompression).
Thus, the second peak represents modes that had time to compress and
then rarefy before the photons were released from the photon-baryon
fluid. The third peak represents
compression-rarefication-compression, and so on.

The height of the first peak is determined by the total energy
density.  Roughly, with more matter, the gravitational attraction
requires more force to counteract, deepening the compression and thus
increasing the amplitude of oscillation.

Now suppose we increase the density of baryons in the photon-baryon
fluid.  This increases the inertia of the fluid, deepening each
compression phase without changing the rarefication. The oscillations
become asymmetric. What this means is the odd-numbered peaks,
whose modes end on a compression, are enhanced relative to the
even-numbered peaks, whose modes end on a rarefication.  Thus as the
baryon fraction increases we should (over some range) see a differential
effect on the odd and even numbered peaks.

The third peak in the spectrum provides the clearest support for the existence
of ``dark matter''  -- a substance of unknown
composition that 
interacts at most weakly 
with baryons (e.g.,
neutrons, protons) or with photons (that's why its dark).
To see why, it is illuminating to compare the oscillations in two example cases.
In the ``radiation-dominated era,'' when photons were the dominant
form of interaction in the universe, density fluctuations were short-lived
and unstable.  A compressed region of photon-baryon fluid would rarefy
as described earlier, but as it did so, the overdensity that caused the original gravitational well would
disappear.  Thus, in this case, at most one cycle of oscillation would
occur between the Big Bang and recombination.  We would see only one
small peak in the temperature power spectrum corresponding to the mode
(component of oscillation) that reaches maximum compression at the
time of recombination.  When the matter fraction is low, the peak
would be small, increasing with the baryon fraction (inertia).

In the ``matter-dominated era,'' however, most of the energy density
was in the form of dark matter.  The
rarefication phase of the oscillation would not eliminate the local
overdensity, allowing multiple cycles of oscillation.  The result is a
spectrum with multiple harmonics and thus multiple peaks.  
The existence and contribution of dark matter is only distinguishable 
from that of baryons alone with three or more peaks.
Moreover, the magnitude of the third peak constrains
the time of transition between a radiation and matter dominated universe.
In particular, a finding that the second and third peaks were comparable in magnitude
would suggest that dark matter dominated before recombination,
which is a fundamental prediction of Big Bang cosmology.
The magnitude of the third peak is also of interest for estimating
the fraction of dark matter in the Universe.
Astronomers have several methods for inferring the dark matter fraction 
(e.g., studying the rotation of galactic disks in the recent Universe), 
and it is vital to determine if these estimates are comparable 
to those produced by the physics of the early Universe.

Finally, the pattern of CMB hot and cold spots we see on the sky corresponds to those photons
just reaching us from the moment of recombination.
(Recombination was relatively quick but not instantaneous, so there is some blurring
of high spatial frequencies from the scatter of photons during that finite period.)
The contribution to this pattern from each acoustic mode maps to a spherical mode of fluctuations
on the sky. The analysis then proceeds by decomposing the observed fluctuations into spherical
modes and using the contributions of these modes to understand the acoustic oscillations.
We discuss this in the next subsection.

%
%
%
%
%
%
%
%
%
%
%
%

\subsection{The CMB Temperature Power Spectrum} \label{sec::spectrum}

Our focus in this paper is inference about the CMB temperature power spectrum
and in particular the peaks in the spectrum.
In this section, we describe the spectrum and some of the issues that arise in estimating it.
Marinucci (2004, this issue) gives a more complete derivation upon which ours is based.

Let $T(\theta,\vartheta)$ denote the temperature field 
as a function of colatitude 
(zero at the zenith) $0\le \theta\le \pi$ and
longitude $0\le\vartheta < 2\pi$.
Let $\overline{T}$ denote the average of the temperature field
over the sphere.

Define
the temperature fluctuation field by
$$
Z(\theta,\vartheta) = \frac{T(\theta,\vartheta) - \overline{T}}{\overline{T}}.
$$
Note that $Z$ is a random field with mean zero
and is assumed to have
finite second moment.
We can expand $Z$
in terms of a orthonormal basis
on the sphere.
The usual choice of basis is the
set of spherical harmonics
$\{Y_{\ell,m}(\theta,\vartheta)\}$,
for positive integers $\ell=1,2, \ldots$
and integers
$-\ell \leq m\leq  \ell$.
(Here $\ell$ is called the multipole index, or loosely ``multipole moment.'')
These are defined as follows:
$$
Y_{\ell,m}(\theta,\vartheta) =
  \sqrt{\left(\frac{ 2\ell + 1}{4\pi}\right)\frac{ (\ell-|m|)!}{ (\ell+|m|)!}}\;
      P_\ell^{|m|}(\cos \theta)\, e^{i m \vartheta},
$$ 
where the $P_{\ell,m}$ $\ell = 1, 2, \ldots$ and $m = 0,\ldots,\ell$ 
are the associated Legendre functions defined by
\begin{eqnarray*}
P_\ell^m(x) &=& (-1)^m (1-x^2)^{m/2} \frac{d^m}{d x^m} P_\ell(x)\\
\noalign{\noindent with Legendre polynomials}
P_{\ell}(x) &=& \frac{1}{2^\ell \ell!} \frac{d^\ell}{d x^\ell}(x^2-1)^\ell.
\end{eqnarray*}

We can now write
\begin{equation}
Z(\theta,\vartheta) = \sum_{\ell=1}^\infty \sum_{m=-\ell}^{\ell}
                           a_{\ell,m}Y_{\ell,m}(\theta,\vartheta) ,
\end{equation}
where,
\begin{equation}
a_{\ell,m} = \int_{0}^{2\pi}\int_{0}^\pi Z(\theta,\vartheta)
Y_{\ell,m}(\theta,\vartheta) \sin \theta d \theta d\vartheta.
\end{equation}
Since $Z$ is a mean zero random field,
the coefficients $a_{\ell,m}$
are random variables.
They have mean 0, variance
$$
C_\ell \equiv \E |a_{\ell,m}|^2,
$$
and are uncorrelated.
The \emph{power spectrum} is defined to be $C_\ell$ as a function of $\ell$.

Usually, it is assumed that
$Z$ is a Gaussian field 
(but see the paper by Marinucci in this issue)
which implies that
the $a_{\ell,m}$ have a Gaussian distribution.
If we were to observe $Z$ without measurement error, 
we could estimate $C_\ell$ by, say,
\begin{equation}\label{eq::Cellhat}
\tilde{C}_\ell = \frac{1}{2\ell +1}\sum_{m=-\ell}^{\ell} a_{\ell,m}^2,
\end{equation}
and thus for large $\ell$ we have $\tilde C_\ell \approx C_\ell$
because we are averaging a large number of $a_{\ell,m}^2$.
We call $\tilde{C}_\ell$ the realized spectrum.
Another important implication of equation (\ref{eq::Cellhat}) is that even with perfect observations,
we would not know the true power spectrum.
Because our Universe is viewed as one realization of a stochastic process,
$\tilde C_\ell$ will in general differ from $C_\ell$, especially for small $\ell$.
This is known as the problem of \emph{cosmic variance}.
We return to this point in Section \ref{sec::discussion}.

In practice, the data are subject to various sources of measurement error, blurring,
and unobserved parts of the sky.
For example,the Milky Way, which is relatively bright,
obscures the deep sky along a wide band.
The spherical harmonics are no longer orthogonal over what is left of the sphere,
which induces correlation and bias into the estimated $C_\ell$s.
There are in addition a host of other complications in measuring $Z$.

Our model, in vector form, is
\begin{equation} \label{eq::linearModel}
\hat{C} = C + \epsilon,
\end{equation}
where $\hat{C}$ is the \emph{observed spectrum} and 
the noise vector $\epsilon$, with covariance matrix $C \aleph C^T$, 
incorporates the known sources of error, including measurement error.
If there were no sky cut for the galaxy, $\aleph$ would
be diagonal, but in practice, it incorporates the various known sources of error.
In practice, the unknown $C$ in the covariance matrix is replaced by a
a pilot estimate, $C^0$.
The choice of $C^0$ turns out to have surprisingly little effect on the results.
We thus take the covariance matrix of $\epsilon$ in equation (\ref{eq::linearModel}) to be
known and equal to $\Sigma = C^0 \aleph (C^0)^T$.

\begin{figure}[t]
\vbox
{
  \hglue -0.5in\psfig{file=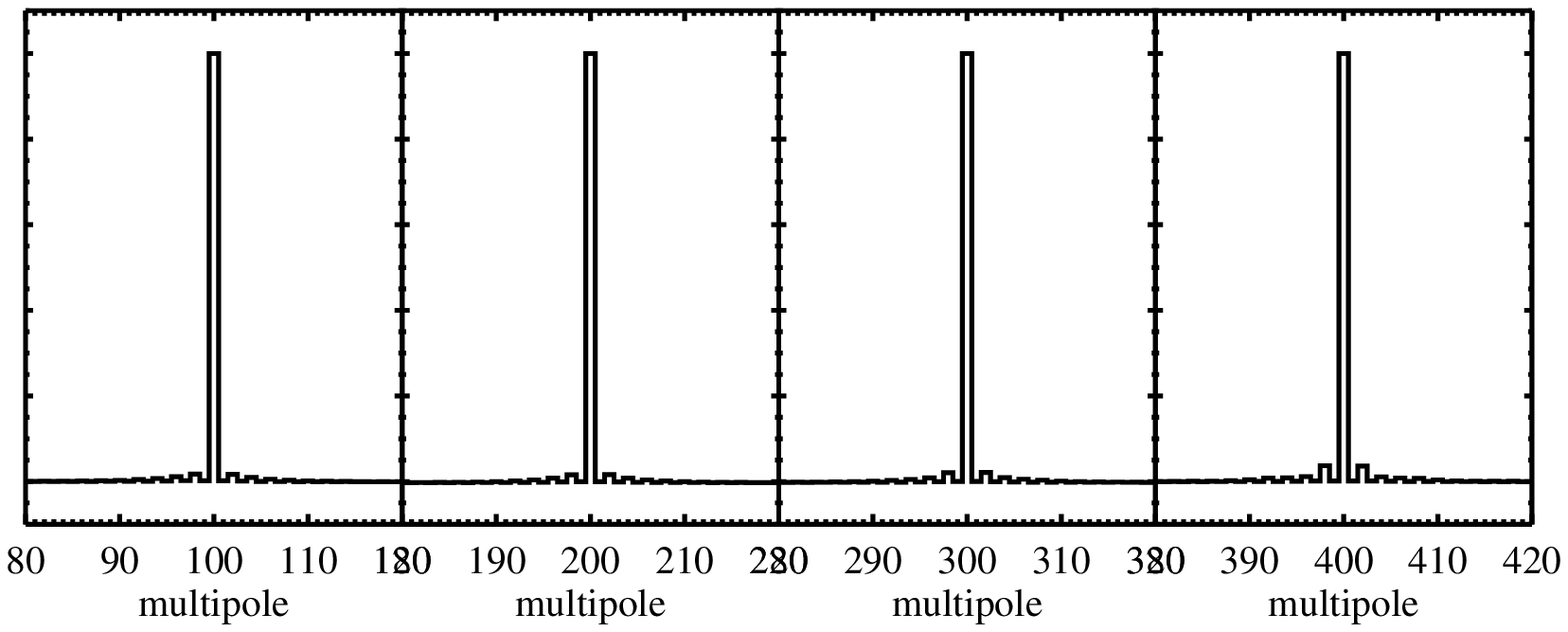,height=2.5in,angle=0}
}
\caption{Bandpower windows from the matrix $K$ centered on (left to right) $\ell=100,200,300,400$.}
\label{fig::window}
\end{figure}

Another issue is that the observations are actually derived from a convolution of the $\tilde C_\ell$s
with $\ell$-dependent window functions; that is, the model is actually $\hat C = K C + \epsilon$
for some matrix $K$.
But as Figure \ref{fig::window} shows, the rows of $K$ are very nearly delta functions. (See Knox, 1999.)
And in fact, incorporating these window functions has negligible effect on our results,
so we disregard them in what follows. 

\section{Uniform Confidence Sets For Nonparametric Regression}

\begin{figure}[t] 
\pspicture(0,0)(16,16)
\rput[bl](-2.5,2.54){\psfig{file=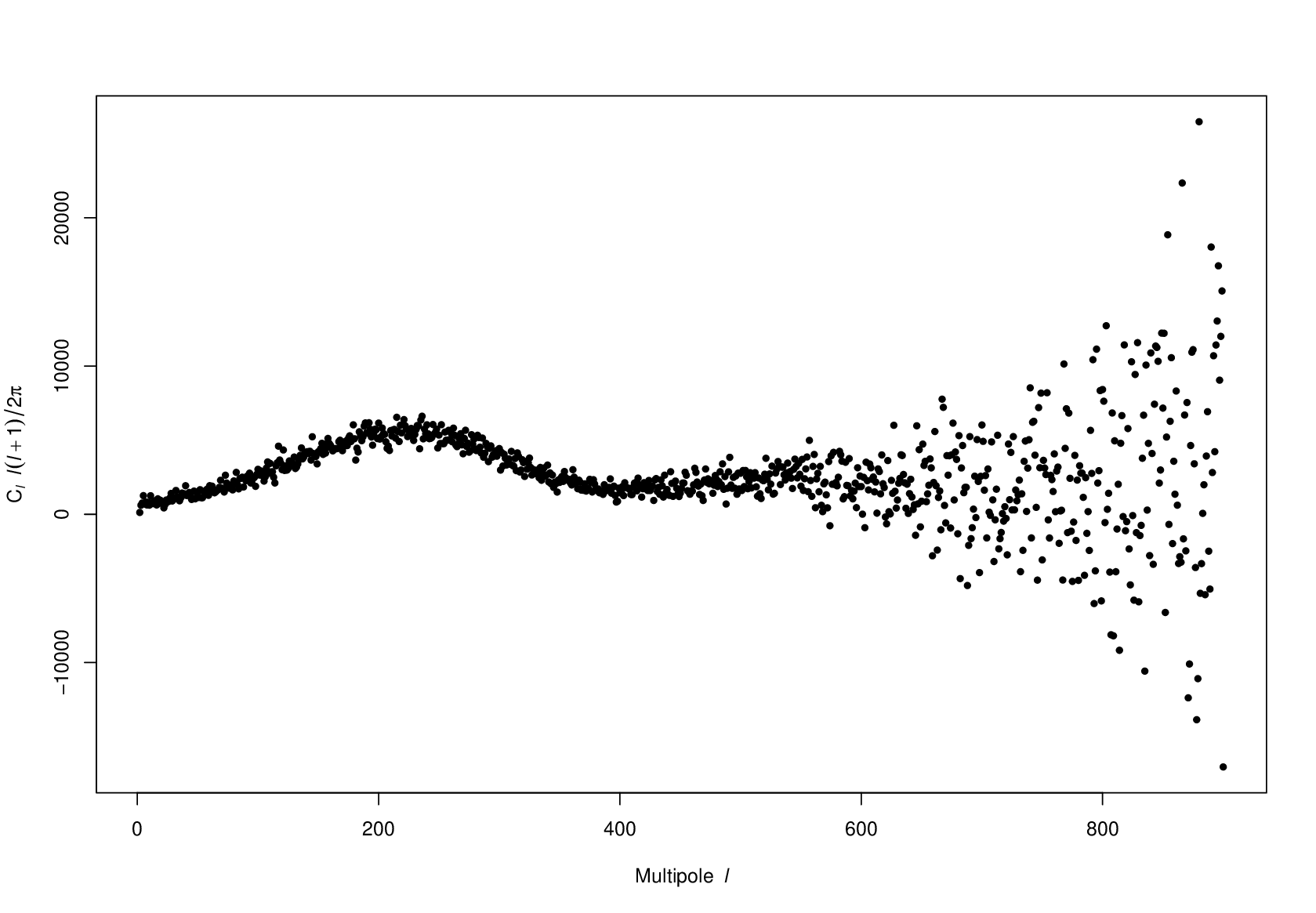,height=5in,angle=270}}
\endpspicture
\medskip
\caption{$Y_\ell$ as a function of $\ell$ for the WMAP data.}
\label{fig::data}
\end{figure}

Taking $Y_\ell = \hat{C}_\ell$ and $x_\ell = \ell/L_{\srm max}$,
let $f(x_\ell)\equiv C_\ell$ denote the true power spectrum
at multipole index $\ell$. See Figure \ref{fig::data} for the $Y_\ell$ from
the WMAP data (Hinshaw et al. 2003).
We can then rewrite equation (\ref{eq::Cellhat}) in the form of a nonparametric regression problem:
\begin{equation}\label{eq::regress2}
Y_\ell = f(x_\ell) + \epsilon_\ell, \qquad \ell = L_{\srm min}, \ldots, L_{\srm max},
\end{equation}
where $\epsilon = (\epsilon_{L_{\srm min}},\ldots,\epsilon_{L_{\srm max}})$ is assumed
Gaussian with known covariance matrix $\Sigma$ as described earlier.
This is only an approximation to the model actually used,
but we will not discuss the various practical complications here.

Let $\sigma_\ell^2$ denote the diagonal elements of $\Sigma$
and $n = L_{\srm max} - L_{\srm min} + 1$ be the total number of observed multipoles.
Henceforth, we will use $i = \ell - L_{\srm min} + 1$ as an index.

Our approach is to nonparametrically estimate the regression
$f$ and find a nonparametric
$1-\alpha$ confidence ball ${\cal B}_n$ for $f$.
More precisely, we want ${\cal B}_n$
\begin{equation}\label{eq::conf-ball}
\liminf_{n\to\infty} \inf_{f\in {\cal F}} P( f \in {\cal B}_n) \geq 1-\alpha
\end{equation}
for some large function class ${\cal F}$
such as a Sobolev space.

Once we have computed the confidence ball, we can construct a confidence interval
for any functional $T(f)$ of interest, such as the location of the first peak.
If ${\cal T}$ is a set of such functionals and
$$
I_n(T)=\left( \min_{f \in {\cal B}_n} T(f),\ \max_{f \in {\cal B}_n} T(f) \right)
$$
then we have that
\begin{equation}
\liminf_{n\to\infty} \inf_{f\in {\cal F}} P\left( T(f) \in I_n(T)\ 
{\rm for\ all\ }T \in {\cal T} \right)  \geq 1-\alpha.
\end{equation}
Alternatively, we can construct the set of cosmological parameters that produces
spectra within the confidence ball, which gives a joint confidence set on these parameters.

We use orthogonal series regression to estimate $f$ and then construct a confidence ball
via the Beran-D\"umbgen pivot method
(Beran 2000, and Beran and D\"{u}mbgen 1998),
which was inspired by an idea in Stein (1981).
Specifically, we expand $f$ in the cosine basis $f = \sum_{j=0}^\infty \mu_j \phi_j$,
where $\phi_0(x) =1$ and $\phi_j(x) = \sqrt{2} \phi(\pi j x)$ for $j\ge 1$.
If $f$ is fairly smooth, for example if $f$ lies in a Sobolev space, then
$\sum_{j>n}\mu_j^2$ is negligible and we can write
$f(x) \approx \sum_{j=0}^n \mu_j \phi_j(x)$.
Let
\begin{equation}
Z_j = \frac{1}{n}\sum_{i=1}^n Y_i \phi_i(x_i)
\end{equation}
for $0\leq j < n$.
Note that vector $Z$ is approximately Normal with mean $\mu$ and variance
matrix $U \Sigma U^T/\sqrt{n}$, where $U$ is the cosine basis transformation matrix.
We define the monotone shrinkage estimator by
\begin{equation}
\hat\mu_j = \lambda_j Z_j
\end{equation}
where
$1 \ge \lambda_1 \ge \cdots \ge \lambda_n \geq 0$
are shrinkage coefficents.
The estimate of $f$ is
$$
\hat{f}(x) = \sum_{j=1}^n \hat\mu_j \phi_j(x).
$$
In this paper, we will use a special case of monotone shrinkage, called nested subset selection (NSS),
in which
$\lambda_j=1$ for $j\leq J$ and $\lambda_j=0$ for $j> J$.
In this case,
$$
\hat{f}(x) = \sum_{j=1}^J Z_j \phi_j(x).
$$

The squared error loss 
as a function of
$\hat\lambda =(\hat\lambda_1, \ldots, \hat\lambda_n)$ is
$$
L_n(\hat\lambda) = \int (\hat{f}(x)-f(x))^2 dx \approx \sum_j (\mu_j - \hat\mu_j)^2.
$$
The risk is
$$
R(\lambda) =\mathbb{E} \int ( f(x) - \hat{f}(x))^2 dx \approx
\sum_{j=1}^n \lambda_j^2 \frac{\sigma_j^2}{n} +
\sum_{j=1}^n (1-\lambda_j)^2 \mu_j^2
$$
where
$\sigma_j^2 = \mathbb{V}(\epsilon_j)$.
The shrinkage parameter $\lambda$ is chosen to minimize
the Stein's unbiased risk estimate
\begin{equation}\label{eq::sure}
\hat{R}(\lambda) =
\sum_{j=1}^n \lambda_j^2 \frac{\hat\sigma_j^2}{n} +
\sum_{j=1}^n (1-\lambda_j)^2  
\left( Z_j^2 - \frac{\hat\sigma_j^2}{n}\right)_+ .
\end{equation}
Beran and D\"{u}mbgen showed that
$\hat{R}(\lambda)$ is asymptotically, uniformally close to
$R(\lambda)$ in either the monotone or NSS case.

The Beran-D\"umbgen method is based on the weak convergence
of the ``pivot process'' 
$B_n(\hat\lambda) = \sqrt{n}(L_n(\hat\lambda) - \hat{R}(\hat\lambda))$
to a Normal $(0,\tau^2)$ for some $\tau^2 > 0$.
(The estimator for $\tau^2$ is given in the Appendix 3.)
It follows that
\begin{eqnarray*}
\cD_n 
  &=& \Set{\mu\st \frac{L_n(\hat\lambda_n)-S_n(\hat\lambda_n)}{\hat\tau_n/\sqrt{n}} \le  z_\alpha} \\
  &=& \Set{\mu\st \sum_{i=1}^n (\hat{\mu}_i - \mu_i)^2 \le 
 \frac{\hat\tau_n \, z_\alpha}{\sqrt{n}} + \hat{R}(\hat\lambda_n)}
\end{eqnarray*}
is an asymptotic $1-\alpha$ confidence set for the coefficients,
where $z_\alpha$ denotes the upper $\alpha$ quantile of a standard Normal
and where $\hat\mu_i \equiv \hat\mu_i(\hat\lambda_n)$.
Thus
\begin{equation}\label{eq::cBn}
\cB_n = \Set{ f(x)=\sum_{j=1}^n \mu_j \phi_j(x):\ \ \mu \in \cD_n }
\end{equation}
is an asymptotic $1 - \alpha$ confidence set for $f$.

The approach to confidence sets that we use here
is quite different than the more familiar
confidence band approach
in which one constructs
bands of the form
$\hat{f}(x) \pm c \sqrt{\hat{{\sf Var}}(\hat{f}(x))}$
for some $c$.
The advantage of bands is that by plotting them,
we get a simple visual impression of the uncertainty.
However, there are some drawbacks to bands.
In their most naive form,
the constant $c = z_{\alpha/2}$,
which does not account for
the multiplicity over the $x$s.
This can be fixed by using a larger constant,
although the computation of the constant is, in some cases, nontrivial.
See Sun and Loader (1994).
Second, the available results about coverage 
appear to be pointwise rather than uniform over $f\in {\cal F}$
although we suspect that the results can be strengthened to be 
asymptotically uniform.
The third, and most serious problem, is that
the function estimate $\hat{f}$ is biased
so the confidence interval is not centered properly, resulting in uncercoverage.
Specifically, letting $s(x)$ denote the standard error of
$\hat{f}$ and $m(x) = \E{\hat{f}(x)}$, we have that
$$
\frac{\hat{f}(x) - f(x)}{s(x)} = 
\frac{\hat{f}(x) - m(x)}{s(x)} + \frac{m(x) - f(x)}{s(x)}.
$$
The first term typically satisfies a central limit theorem.
The second term does not tend to zero since
optimal smoothing causes the bias 
$m(x) - f(x)$ to be of the same order as $s(x)$.
There have been some attempts to control this
smoothing bias; see Ruppert, Wand and Carroll (2003) for 
a discussion.

The confidence ball approach automatically deals with the smoothing bias,
at least approximately.
This is because the ball takes the object
$||\hat{f}(x) - f(x)||^2$ as its starting point, rather than
$\hat{f}(x) - m(x)$ which is implicit in the band approach.
The ball approach does have some bias,
since $\hat{f}$ actually estimates
$f_n(x) = \sum_{j=1}^n \mu_j \phi_j(x)$ rather than
$f(x) = \sum_{j=1}^\infty \mu_j \phi_j(x)$ 
resulting in a tail bias of $\sum_{j=n+1}^\infty \mu_j^2$.
However, this tail bias is small relative to the smoothing bias.

\section{Dealing with Heteroskedastic Errors} \label{sec::hetero}

\begin{figure}[t] 
\pspicture(0,0)(16,16)
\rput[bl](-2.5,8.7){\psfig{file=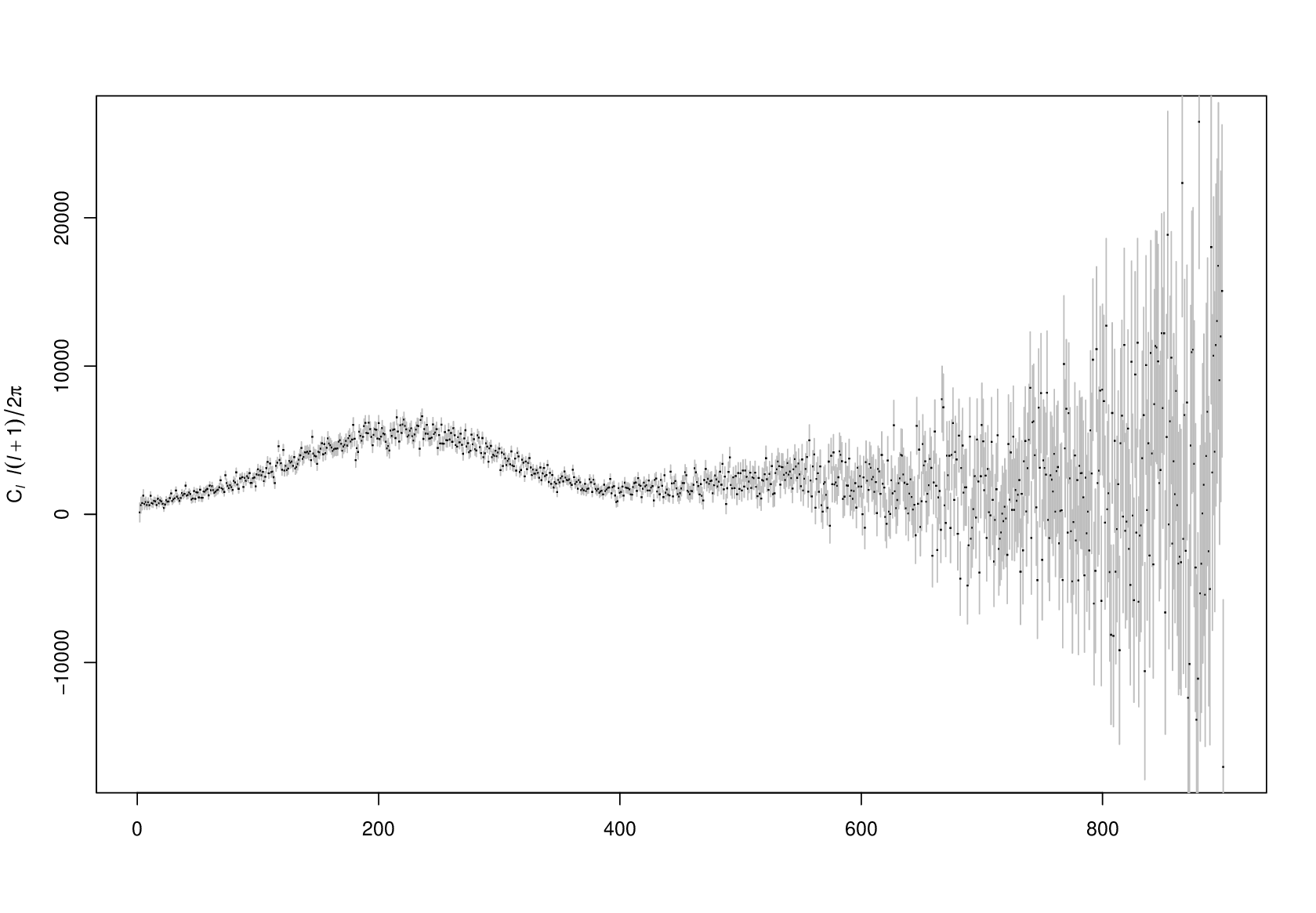,height=5in,angle=270}}
\rput[bl](-2.5,-1){\psfig{file=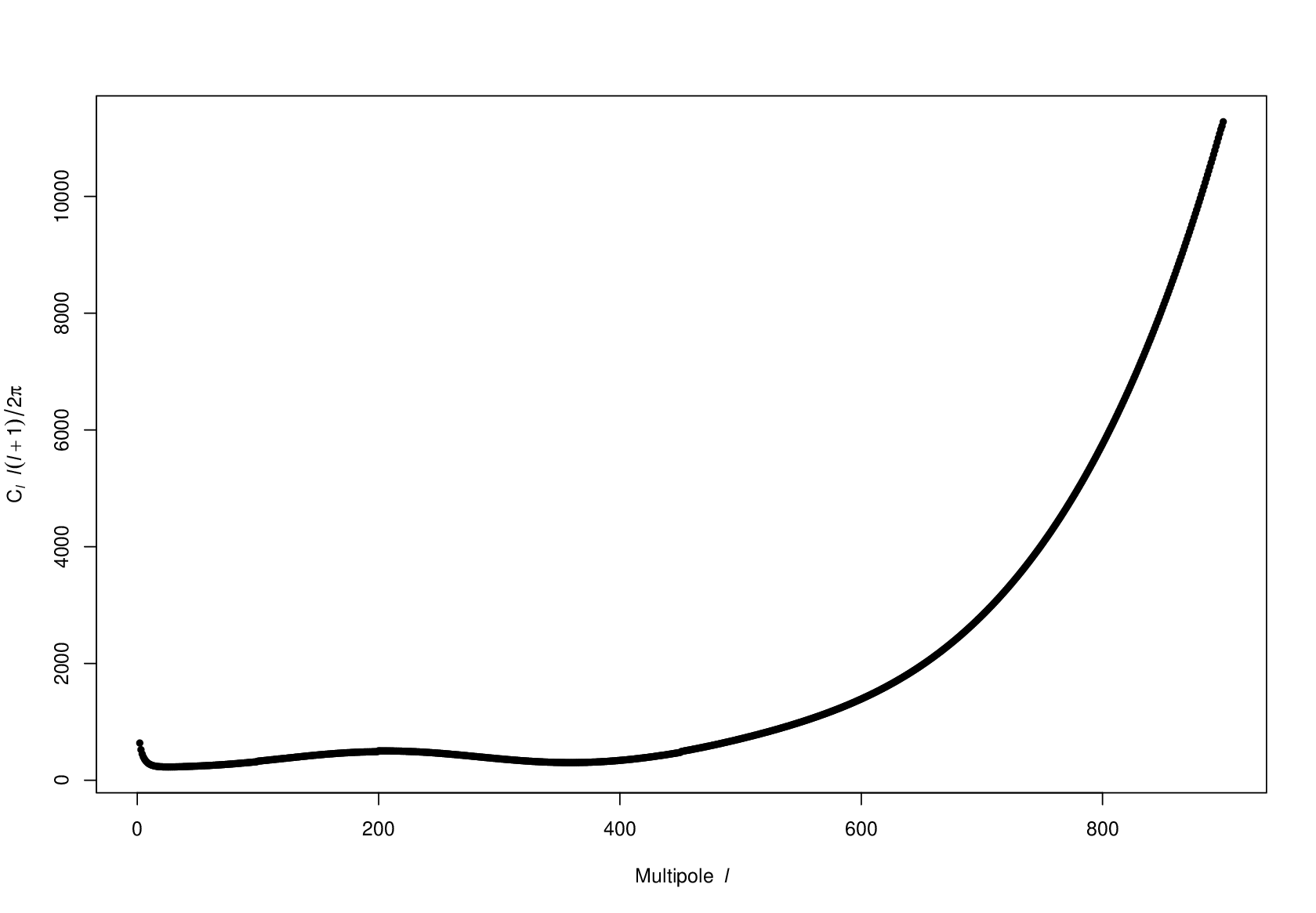,height=5in,angle=270}}
\endpspicture
\medskip
\caption{Noise standard deviation as error bars on data (above)
and as a function of $\ell$ (below).}
\label{fig::variances}
\end{figure}

As Figure \ref{fig::variances} shows,
the data for the CMB power spectrum are highly heteroskedastic.
The confidence set based on $\cL^2$ loss is a ball
and thus gives equal weight to deviations in all direction.
Because the 
the CMB variances are tiny for some $\ell$s and huge for others,
this symmetry is inappropriate.
In parametric inference, confidence sets under heteroskedasticity are
typically ellipses rather than balls, and we need to make a similar
adjustment.
We do this by constructing the confidence set under
a loss function that gives more weight to points where the spectrum
is measured precisely.
In this section, we extend the Beran-D\"umbgen method to such weighted loss functions.

We now replace the $L^2$ loss function with the
following weighted loss:
$$ L(f,\hat{f}) = \int (f - \hat{f})^2 w^2,$$
where we take $w^2(x) = 1/\sigma^2(x)$.
We expand both the unknown function and the weight function $w^2$ in the
orthonormal basis.
Hence, we write
\begin{eqnarray*}
  f(x) &=& \sum_j \beta_j \phi_j(x)\\
w^2(x) &=& \sum_j     w_j \phi_j(x),
\end{eqnarray*}
where $\phi_0, \phi_1, \ldots$ is the cosine basis on [0,1] defined above.

The construction of $\cB_n$ requires a new central limit theorem and a modified estimate
of the asymptotic variance.
We also replace the risk estimator in equation (\ref{eq::sure}) by
the following, which can be shown to be unbiased for the new loss function:
\begin{equation} \label{eq::Rhat}
\hat R = Z^T \bar D W \bar D Z + \trace(D W D B) - \trace(\bar D W \bar D B),
\end{equation}
where $D$ and $\bar D = I - D$ are diagonal matrices with 1's in the first $J$ and last $n - J$
entries,
$B = U \Sigma U^T$ is the covariance of $Z$,
and $W_{jk} = \sum_\ell w_\ell \Delta_{jk\ell}$
with $w_\ell$ being the $\ell^{\rm th}$ expansion coefficient of the function $w^2$
and
\begin{eqnarray*}
\Delta_{jk\ell}
 &=& \int_0^1 \phi_j\phi_k\phi_\ell \\
 &=& \cases{ 
   1 & if $\#\{j,k,\ell = 0\} = 3$\cr
   0 & if $\#\{j,k,\ell = 0\} = 2$\cr
   \delta_{jk}\delta_{0\ell} + \delta_{jl}\delta_{0k} + \delta_{kl}\delta_{0j}& if $\#\{j,k,\ell = 0\} = 1$\cr
   \frac{1}{\sqrt{2}} (\delta_{\ell,j+k} + \delta_{\ell,|j-k|}) & if $j,k,\ell > 0$. \cr }
\end{eqnarray*}
The set $\cB_n$ is defined as in equation (\ref{eq::cBn})
but with the new estimate of risk.
The estimated variance of the pivot, $\hat\tau^2$, is also different and is given in
the appendix.

\section{Results}\label{sec::results}

\begin{figure}[t] 
\hbox{\hskip -1in\psfig{file=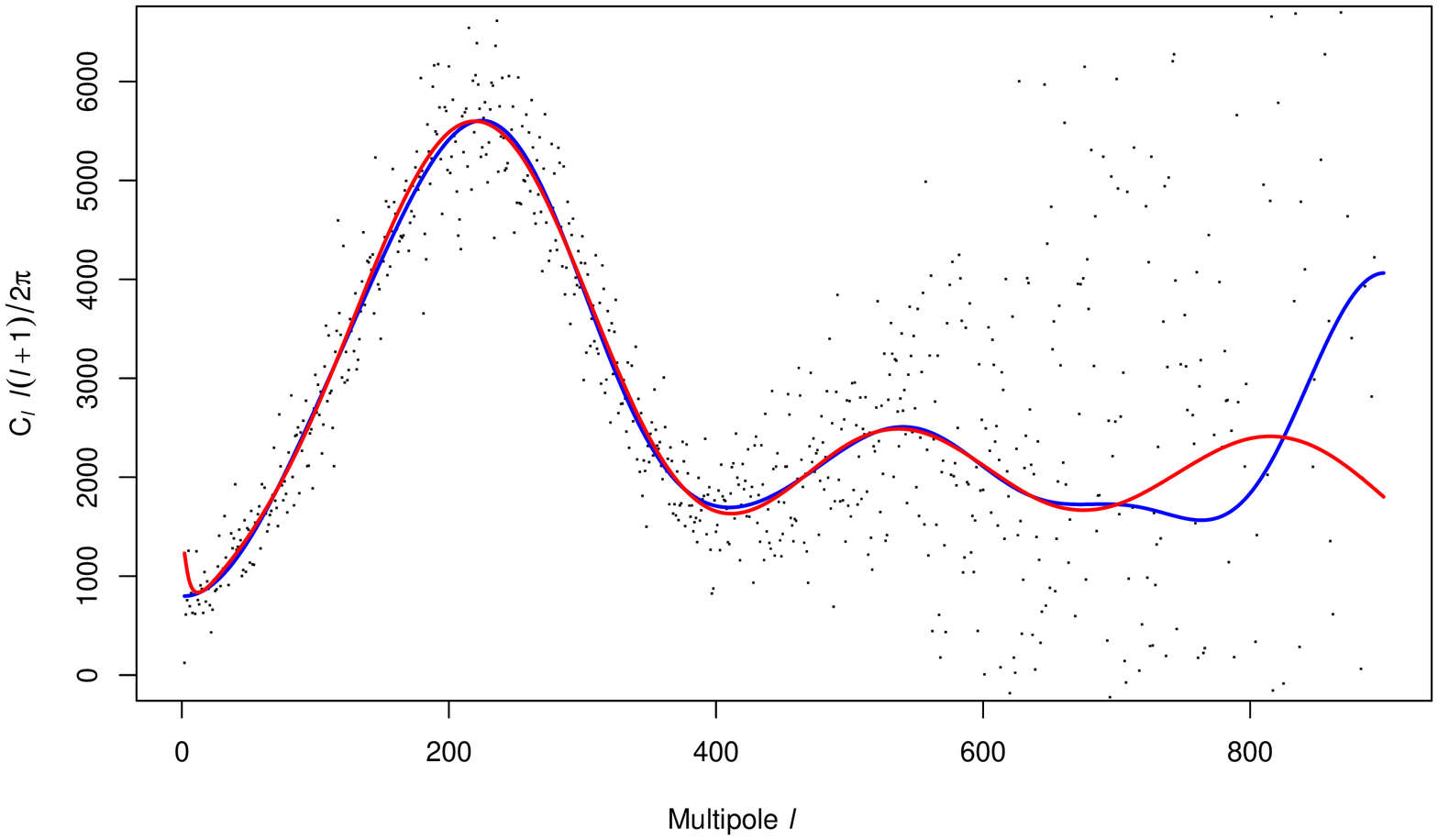,height=5in,angle=0}}
\caption{Center of our confidence ball 
(curve with sharp rise at right)
and the power spectrum
for the Concordance model 
(curve with three peaks).
Note the striking agreement between the nonparametric fit
and the parametric fit.}
\label{fig::wmapFit}
\end{figure}

\begin{figure}[t]
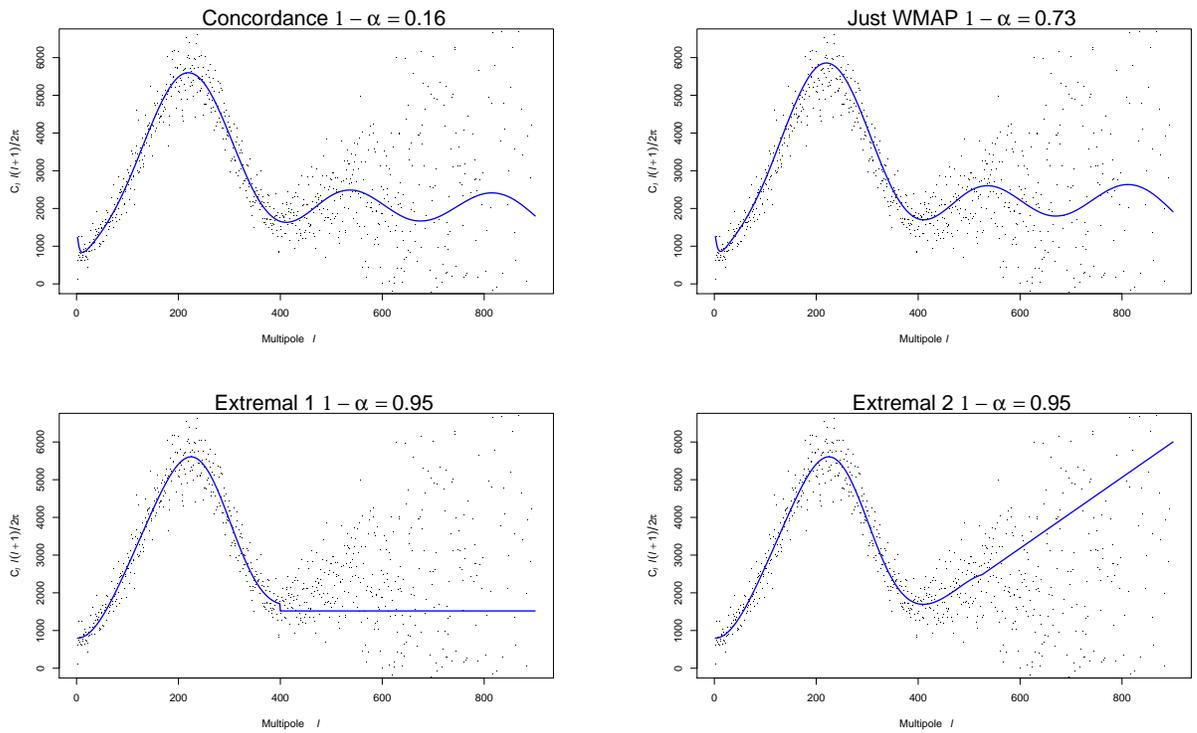

\def\figcell#1{\hfil\hbox{\hskip -1em\psfig{file=#1.eps,height=2in,angle=0}}\hfil}
$$
\halign{\tabskip=0in\figcell{#}\tabskip=0.5in&\figcell{#}\tabskip=0em\cr
wmap-concord2&wmap-just\cr
wmap-ext1&wmap-ext2\cr
}
$$
\caption{CMB spectra: (top left) Concordance model fit, (top right) WMAP-only model fit,
and (bottom) two extremal fits.}
\label{fig::wmapExtremes}\refstepcounter{figure}
\end{figure}

We applied our method to the WMAP data to obtain a confidence
set for the unknown spectrum $f(\ell/L_{\srm max}) \equiv C_\ell$.
Figure \ref{fig::wmapFit} compares the center of our confidence ball
with the so-called ``Concordance model'' (Spergel et al. 2003).
The Concordance model is the maximum likelihood estimator for a likelihood
of the form 
\def\ssrm{\scriptscriptstyle\rm}
\begin{eqnarray} 
\cL_{\ssrm Conc}(\theta;Y_{\ssrm WMAP}, Y_{\ssrm LSS}, Y_{\ssrm Lyman}, Y_{\ssrm CBI}, Y_{\ssrm Acbar}) 
\hskip -2.6in && \nonumber\\
  &=& \cL_{\ssrm WMAP}(\theta;Y_{\ssrm WMAP})\cdot \nonumber\\
  && \cL_{\ssrm LSS}(\theta;Y_{\ssrm LSS}) \cdot \cL_{\ssrm Lyman}(\theta;Y_{\ssrm Lyman}) \cdot \cL_{\ssrm CBI}(\theta;Y_{\ssrm CBI}) \cdot \cL_{\ssrm Acbar}(\theta;Y_{\ssrm Acbar}), \label{eq::concord_like}
\end{eqnarray}
where the $Y$s are independent data sets from different experiments
(WMAP: Bennett et al. 2003; LSS: Percival et al. 2003; Lyman: Croft et al. 2002, Gnedin and Hamilton 2002; CBI: Mason et al. 2003, Sievers et al. 2003, Pearson et al. 2003; Acbar: Kuo et al. 2001).
In particular, $Y_{\ssrm WMAP}$ is the data set we are using.
The parametric fit from the WMAP data alone (see Figure \ref{fig::wmapExtremes}, top right)
is obtained by maximizing only the first component $\cL_{\ssrm WMAP}(\theta;Y_{\ssrm WMAP})$.

Note how well the nonparametric curve compares to the Concordance spectrum.
The notable exceptions are in the very high-$\ell$ region around the third peak
and the low-$\ell$ region where the physical models curve upward sharply.
We will argue that both the third peak and the rise in the spectrum at low $\ell$s 
are by-products of the model and not the data.
All of the cosmological models share both features.
We are not suggesting that these features are incorrect,
but we believe it is useful to separate effects driven by the data from those driven by the model.
See Section \ref{sec::discussion}.

Once we construct the confidence ball, the next step is to use it
to draw inferences.
Because the ball is 900-dimensional in this case, it can seem daunting
to display results.
Fortunately, our construction provides simultaneous coverage over all functionals
of the unknown function, pre or post hoc.
We thus explore the uncertainty by creating targeted \emph{probes} of the ball.

\begin{figure}
\hbox{\hskip -1in\psfig{file=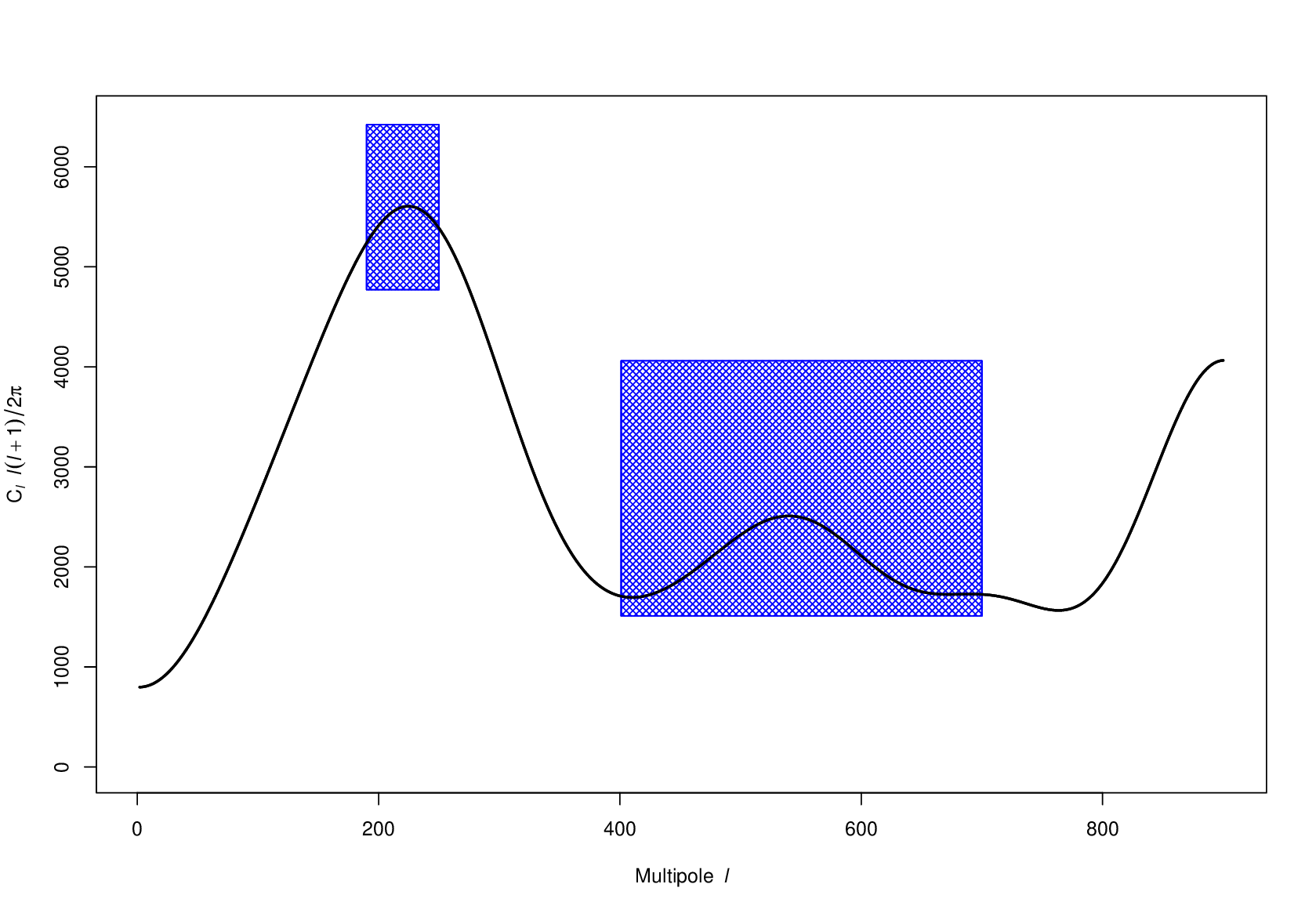,height=5in,angle=270}}
\caption{Center of our 95\% confidence ball with superimposed 
95\% intervals for the heights and widths of the first two peaks.}
\label{fig::wmapFpeakProbe}
\end{figure}

First, we explore the uncertainty in the location and height of the peaks.
To do so, we searched through the confidence ball using
local quadratic probe. Specifically,
at each location $\ell_0$,
we defined a quadratic $q_h(\ell)$ with support on the interval $[\ell_0 - \Delta,\ell_0 + \Delta]$
for fixed $\Delta = 51$, centered at $\ell_0$, and with height $h$.
If $\hat f$ is the center of our confidence ellipse,
we considered perturbations of the form $f = \hat f + q_h$.
We varied $h$ to find the largest and smallest values
such that the resulting $f$ is within the confidence ball and
maintains three peaks over the $\ell$ range $[2,900]$.
This results in confidence limits on the peaks as shown in Figure \ref{fig::wmapFpeakProbe}.

One striking result is the different precisions with which
the first and second peaks are resolved.
This is to be expected given the large variances
near the second peak.
In other words, the data alone give little information about
the second peak. (The third peak is even more uncertain.)
The published results in the physics literature 
present the second peak with much lower uncertainty.
We return to this issue in Section \ref{sec::discussion}.

\begin{figure}[t]
\hbox{\hskip -1in\psfig{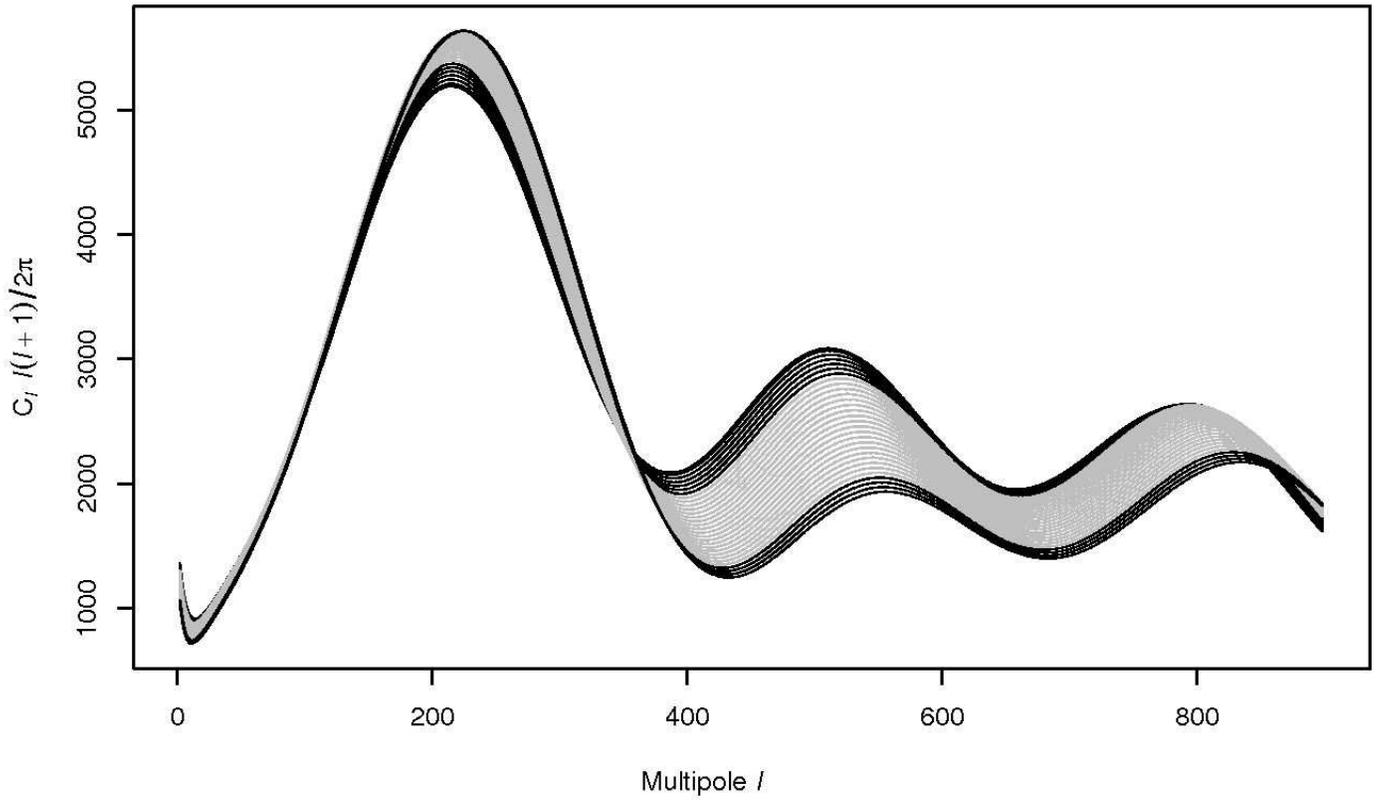}}
\caption{``Ribbon'' probe of the confidence ball within the parametric
model keeping $\Omega_{\srm Total}$ fixed at 1 and varying the Baryon fraction $\Omega_{\srm b}h^2$
from 0.01225 to 0.03675.}
\label{fig::wmapFProbe}
\end{figure}

Figure \ref{fig::wmapFProbe} shows an example of a model-directed probe.
Using the {\sc CMBfast} software package (Seljak and Zaldarriaga 1996),
we generated spectra in a one-dimensional family centered on the Concordance model parameters.
The figure, which we call a ribbon plot,  
shows how the spectrum changes as the baryon fraction $\Omega_{\srm b}$ 
is varied while keeping the total energy density $\Omega_{\srm Total}$ fixed at 1.
The light gray curves are in the ball; the black curves are not.
The resulting interval for $\Omega_b h^2$ is [0.0169,0.0287].
To generate a valid confidence interval with such a probe 
we would need to search the entire 11-dimensional parameter space.

\begin{figure}[t]
\hbox{\hskip -1in\psfig{file=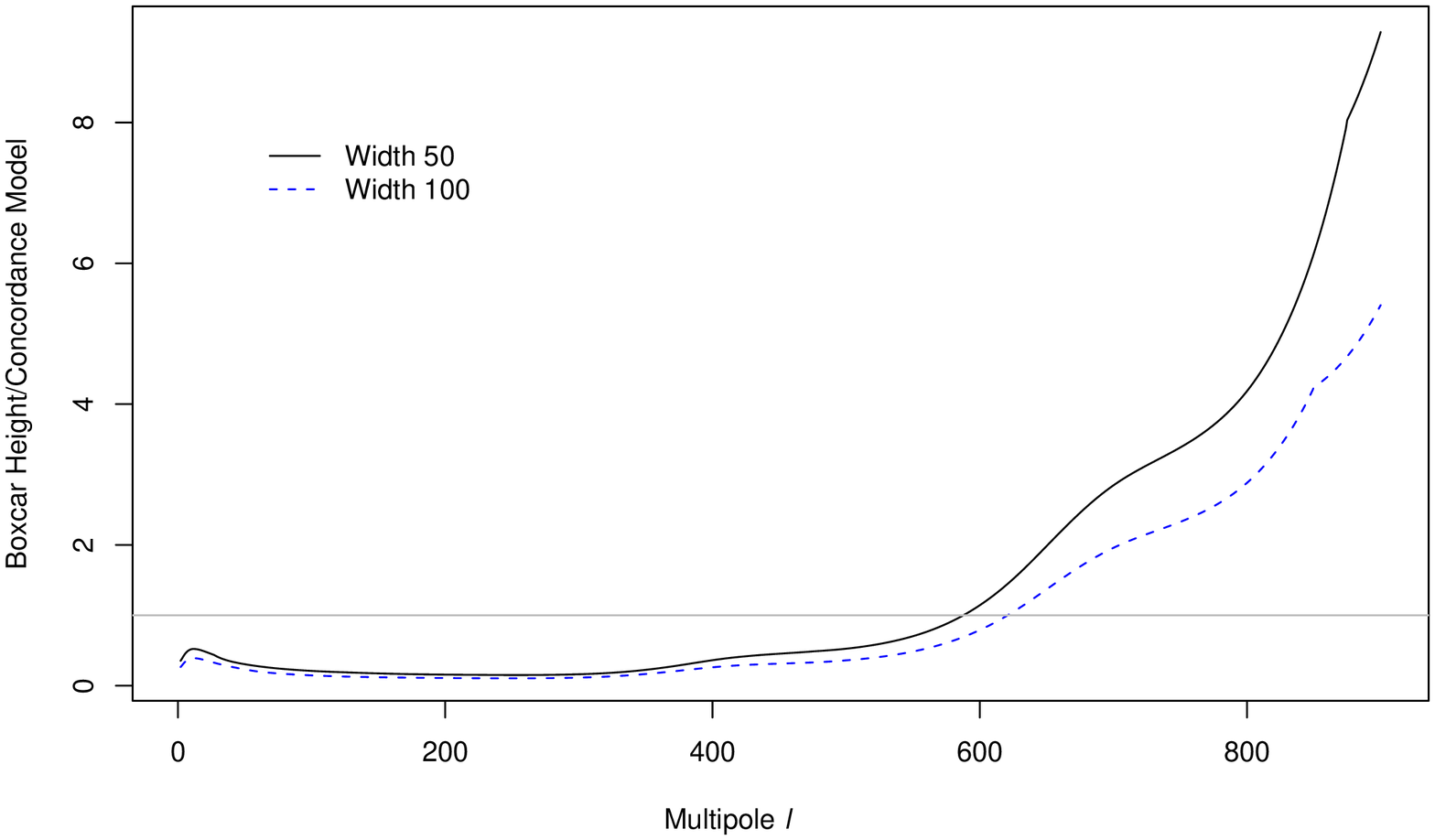,height=5in,angle=0}}
\caption{Height of local ``box car probe'' that is just in the 95\%
confidence ball, divided by the height of the Concordance spectrum, for two
different box car widths. The horizontal line is at height 1.}
\label{fig::wmapBoxcar}
\end{figure}

The data are much noisier for high $\ell$s, 
and we want to quantify how this propogates into local
uncertainty about the function 
because this our ability to resolve the second and third peaks.
A simple probe of the confidence set is useful for this purpose:
finding how far a particular function in the confidence ball 
can be perturbed by localized deviations.
For example, 
at each $\ell$,
we examined the one-dimensional family of spectra 
$f_h = \hat f + h\cdot b$,
where $b$ is a boxcar of fixed width and unit height centered at $\ell$.
Figure \ref{fig::wmapBoxcar} shows the maximum absolute height $h$ that remains in
the 95\% ball relative to the height of the Concordance spectrum, for two different boxcar widths.
At $\ell$s where this curve is greater than 1, 
the data arguably contain little information about
the height of the curve near that location.

The confidence ball is also useful for model checking. 
Figure \ref{fig::wmapExtremes} shows four different spectra
along with the minimum value of $1 - \alpha$ for which each spectrum
is in the $1-\alpha$ confidence ball.
The concordance spectrum is very close to the center.
But the best fitting parametric model using only the WMAP data is at best
in the 73\% confidence ball.
Cosmologists often use 68\% confidence levels, so this can be seen as a weak
rejection of the best fitting model from the data.
We also considered two extremal models that are in the 95\% ball.
These show that the data alone are consistent with eliminating the second
and third peaks.
While the cosmological models all predict these peaks -- through the acoustic oscillations
caused by dark matter -- this suggests the benefits of more precise data,
as from the Planck mission (Balbi et al. 2003).

\section{Discussion}\label{sec::discussion}

\subsection{Findings}

Our most striking finding is that 
the center of our nonparametric confidence ball using the WMAP data alone
lies very close to the Concordance model fit
over the range where the data are not noise dominated.
Recall from equation (\ref{eq::concord_like})
that the Concordance model incorporates four independent data sets,
each with distinct likelihood forms.
In contrast, the parametric model fit using only the
WMAP data (with likelihood $\cL_{\ssrm WMAP}$) 
lies barely in the 73\% confidence ball.  
Given that cosmologists often use 68\% confidence intervals as their standard of
evidence, this is tantamount to a rejection of the cosmological model
that underlies that parametric fit.  

This raises two points.  
First, it is remarkable that with a fully nonparametric method 
we have come very close to the Concordance model
based on the WMAP data alone.  Second, that we obtained basically the same
spectrum as the Concordance model calls into question
the accuracy of the WMAP-only likelihood $\cL_{\ssrm WMAP}$.

\subsection{Methods}

We have presented a nonparametric method for analyzing the CMB
spectrum.  Our techniques have wide applicability to regression
problems beyond cosmology.  By starting with a confidence ball, then
probing the ball using functionals, one can address a variety of
questions about the unknown function while maintaining correct
coverage, despite multiplicity and post-hoc selection.

The method in this paper modifies the original Beran-D\"umbgen construction
to account for heteroskedasticity.
This modification yields a substantial reduction in the size of the confidence set.
The resulting confidence set is also more useful in that 
it leads to tighter (looser) bounds in regions where the function
is more (less) accurately measured.

One advantage of our approach, is that it allows one to separate the
information in the data from the information in a model.  Although
we did not pursue the full calculation here, we could intersect our
confidence ball with the manifold of spectra from the parametric model as a way
to combine data and model.  Specifically, we could use the cosmological model to
generate spectra, but then test which spectra are consistent with the
data by reference to our confidence ball.  This does not rely on
likelihood asymptotics which, as we discuss below, are suspect in this
problem.  Another advantage is that by extending this analysis to a
constrained noparametric model (such as a three peak model) that
contains the cosmological model, we can make the same inferences
without being tied to the analytic form of the model.
Our approach can then be used to 
check the model, make inferences under the model, and
compare parametric to nonparametric inferences.

We should point out that cosmologists obtain confidence intervals for
parameters in their (11-dimensional) model by integrating over the
nusiance parameters and producing a marginal posterior.  However, the
likelihood is ill-behaved, under-identified, and degenerate.  Moreover, in
the physics literature, there does not seem to be a clear appreciation
of the fact that interval estimates obtained this way need not have
correct frequentist coverage.

There are several other advantages to our approach.  If a parameter
is under-identified this will show up automatically as a
wide confidence interval.  The intervals have correct asymptotic
coverage and simultaneous validity over all parameters of interest.  There is no
need to integrate or profile the likelihood function.  Finally, the
asymptotic theory for (\ref{eq::conf-ball}) is insensitive to the fact
that the standard asymptotics for the likelihood approach fail.

\subsection{Inferential Foundations}

Interestingly, there seems to be some confusion about the validity of
frequentist inference in cosmology.  Since we have access to only one
Universe -- and thus cannot replicated it --  some feel that it makes no sense to make frequentist
inferences.  This represents a common misunderstanding about frequentist inference in general
and confidence intervals in particular.  The frequency statements for
confidence intervals refer to the procedure, not the target of the
inference.  Our method for constructing confidence balls traps the
true function 95 percent of the time, even over a sequence of different,
unrelated problems. There is no need to replicate the given experiment,
or Universe.  

Complicating matters is the fact that the coefficients $a_{\ell,m}$ of the
temperature field are random and unknown.  To see the importance of this
point, it is useful to make a finer distinction by defining the
realized spectrum $\tilde{C}_\ell = (1/(2\ell+1))\sum_\ell |
a_{\ell,m}|^2$, the ``true spectrum'' $C_\ell=\E(\tilde{C}_\ell )$ and
the measured spectrum $\hat{C}_\ell$.  Note that all our inferences
have actually been directed at the realized spectrum.  Some phyciststs
find it disturbing to be making frequentist inferences about
$\tilde{C}_\ell$ since it is a realization of a random variable rather
than a parameter in the usual sense.  But this is no different than
making inferences about a random effect in a standard random effects
model.

These confusions have led to an interesting movement towards Bayesian
methods in cosmology.  Of course, when used properly, Bayesian methods
can be very effective. Currently however, the Bayesian interval estimates in
the physics literature seem questionable, 
being based on unfettered use of marginalizing over high-dimensional, degenerate likelihoods
using flat priors chosen mainly for convenience.  Indeed, an active
area of research is finding corrections for such intervals to make
them have correct coverage.  Moreover, the potentially poor coverage
of the Bayesian interval seems not to have been widely recognized in
the Physics literature.

\section*{Appendix 1. CMB Data.}

The CMB is composed of photons. The temperature of these
photons ($2.726$ Kelvin) means that the radiation will be at the microwave
wavelengths. The light is collected via a dish (or reflector) and fed into
either a (1) bolometer, which senses small temperature change as the
photon hits the detector, or (2) a high performance transistor. In some
cases (such as the aforementioned COBE experiment), the telescope is
placed in orbit above the Earth. In other cases, the telescope is placed
on a balloon and launched into the atmosphere. With careful attention paid
to ground reflections, CMB telescopes
can also be placed on the ground in regions where the atmosphere will contribute
little contamination (like the South Pole). In all cases, there are a series
of steps leading from the raw data collection 
to the final power spectrum
estimation. 

The raw data are collected in a time stream.  For each moment in time, the
telescope records a temperature difference on the sky between two
widely separated points. For example, one of these locations could
be a fixed source of known temperature, thus allowing the
temperature at the other point to be calculated. However, the comparison
location need not be fixed (or known) and the absolute temperatures
can still be solved for iteratively,
using previous measurements. 
Throughout this process, the {\it pointing}
of the telescope needs to be accurately known (as a function of time), as well as
the {\it calibration} of the temperatures, and also the instrument noise.

\section*{Appendix 2. Cosmological Parameters}

The physics of the Universe on large scales is well described by a small set of
cosmological parameters.
We describe some of the most important parameters below.

\emph{Normalized Hubble Constant $h$}. The Hubble constant is the rate of the
Universe's expansion. Specifically, $H = \frac{\dot{a}}{a}$ where $a$
is the size of the Universe and $\dot{a}$ is the rate of change in $a$.
``Constant'' is a misnomer since this is a dynamic quantity.
The ``Hubble constant'' refers to the value of $H$ as measured today ($H_0$);
this is often normalized and reported as $h$ = $H_0/100.0$ with units km s$^{-1}$ Mpc$^{-1}$.

\emph{Total Energy Density $\Omega_{\srm Total}$}. 
$\Omega_{\srm Total}$ is the energy density of the Universe divided by
the critical density of the Universe: $\rho_{crit} = 3c^2H_0^2/8\pi G$ 
at which the Universe would be geometrically flat.
$\Omega_{\srm Total}$ can be broken down into the sum of different components, like
$\Omega_{\rm{baryons}}$, $\Omega_{\rm{dark~matter}}$, $\Omega_{\rm{neutrinos}}$.

\emph{Cosmological Constant $\Lambda$}. $\Lambda$ is a constant that was added by Einstein into his
general relativistic field equations to produce a static Universe. The constant was later dismissed
as unnecessary after the discovery by Edwin Hubble that the Universe is not static, but expanding. 
However, recent studies show strong evidence for a cosmological constant term. $\Lambda$ acts
as a negative pressure and thus might accelerate the expansion of the Universe. We
often speak of the energy density component $\Omega_{\Lambda}$, which is then
included in the sum of $\Omega_{\rm{Total}}$.

\emph{Baryon Density $\Omega_{\srm b}$}. This is the density component of baryonic matter in the
Universe (e.g. protons, neutrons, etc). The fraction of 
matter density that is baryonic (over the total matter density of the Universe
which includes baryons and non-baryonic dark matter) is often measured to be in the range:
15\%-20\%.

\emph{Dark Matter Density $\Omega_{\srm d}$}. The majority of matter in the Universe is
detected indirectly through it's gravitational effects. Since it cannot be seen or measured
in the laboratory, it has been dubbed ``dark matter''. $\Omega_{\srm d}$ is the energy density
component strictly due to dark matter.

\emph{Neutrino Fraction, $f_{\nu}$}. The fraction of the neutrino density over the total matter
density: $f_{\nu} = \Omega_{\nu}/(\Omega_{\srm b} + \Omega_{\srm d})$.

\emph{Optical Depth $\tau$}. We know today that most of the hydrogen in the Universe is ionized. So at
some time after recombination,
the Universe was re-ionized. $\tau$ is the optical depth due
to Thomson scattering up to a 
redshift of $z < z_{ionization}$: $\int_0^{t(z_{ionization})} \sigma_T n_e dt$
where $\sigma_T$ is the Thomson scattering cross-section and $n_e$ is the electron density.

\emph{Spectral Index $n_s$}.  The primordial matter density fluctuation spectrum is
proportional to the scale size raised to the power $n$, 
the primordial spectral index. On large enough scales, the
the CMB temperature power spectrum's spectral index ($n_s$) is then close (or equal) to the primordial
spectral index.

The spectrum may be approximated numerically as a function of these parameters
using the {\sc CMBfast} software package
(Seljak and Zaldarriaga 1996). Figure \ref{fig::wmapFProbe} shows spectra
corresponding to a range of cosmological parameter settings. 
For example, the location
and amplitude of the first peak is related to the total energy density 
$\Omega_{\srm Total}$. 
The baryon fraction $\Omega_{\srm b}$ and the spectral index $n_s$
drive the ratio of the amplitude of the first and second peaks. 
The ratio of the amplitudes of the second to
third peaks depends on the density of matter 
$(\Omega_{\srm b} + \Omega_{\srm d} + f_\nu)$, $h$, and $n_s$.

\section*{Appendix 3. Estimating $\tau$.}


Recall from Section \ref{sec::hetero} that the cosine basis is defined on [0,1] by
$$
 \phi_0(x) = 1,\ \ \  \phi_j(x) = \sqrt{2}\cos(\pi j x) \quad j > 0.
$$
Then, if $j$ and $k$ are distinct and positive,
\begin{eqnarray*}
  \phi_j \phi_k &=& 2 \cos(\pi j x) \cos(\pi k x) \\
                &=& \cos(\pi (j+k) x) + \cos( pi (j-k) x) \\
                &=& \frac{1}{\sqrt{2}}\left( \phi_{j+k} + \phi_{|j-k|} \right).
\end{eqnarray*}
If $j > 0$
$$
  \phi_j^2  = 2 \cos^2(\pi j x) 
            = \cos(\pi 2 j x) + 1 
            = \frac{1}{\sqrt{2}}\phi_{2j} + \phi_0.
$$

Hence,
\begin{eqnarray*}
\Delta_{jk\ell}
 &=& \int_0^1 \phi_j\phi_k\phi_\ell \\
 &=& \cases{ 
   1 & if $\#\{j,k,\ell = 0\} = 3$\cr
   0 & if $\#\{j,k,\ell = 0\} = 2$\cr
   \delta_{jk}\delta_{0\ell} + \delta_{jl}\delta_{0k} + \delta_{kl}\delta_{0j}& if $\#\{j,k,\ell = 0\} = 1$\cr
   \frac{1}{\sqrt{2}} (\delta_{\ell,j+k} + \delta_{\ell,|j-k|}) & if $j,k,\ell > 0$ \cr }
\end{eqnarray*}
We thus have that
\begin{eqnarray*}
L(f,\hat{f}) &=& \int (f - \hat{f})^2 w^2 \\
             &=& \sum_{j,k,\ell} (\beta_j - \hat\beta_j)(\beta_k - \hat\beta_k) w_\ell 
                                 \int \phi_j \phi_k \phi_\ell  \\
             &=& \sum_{j,k} (\beta_j - \hat\beta_j)(\beta_k - \hat\beta_k) \sum_\ell w_\ell \Delta_{jk\ell} \\
             &=& (\beta - \hat\beta)^T W (\beta - \hat\beta),
\end{eqnarray*}
where $W_{jk} = \sum_\ell w_\ell \Delta_{jk\ell}$.

Let $\bar\lambda = 1 - \lambda$ and let $D(x)$ denote the diagonal matrix
with $x$ along the diagonal.
Write $\hat\beta = D(\lambda) Z$.
Assume $Z$ has a Normal$\langle \beta, B\rangle$ distribution.
Then, $\E\hat\beta = D(\lambda) \beta$
and since $\Cov(\hat\beta_j,\hat\beta_k) = \lambda_j \lambda_k B_{jk}$,
$\Var(\hat\beta) = D(\lambda) B D(\lambda)$.
Then,
\begin{eqnarray*}
  \E L
     &=& \E (\hat\beta - \beta)^T W (\hat\beta - \beta) \\
     &=& \trace(D(\lambda) W D(\lambda) B) + \beta^T D(\bar\lambda) W D(\bar\lambda) \beta.
\end{eqnarray*}
The latter quadratic form can be written as $\sum_{j,k} \beta_j\beta_k \bar\lambda_j\bar\lambda_k W_{jk}$.
We obtain unbiased estimate $\hat L$ by replacing $\beta_j\beta_k$ with $Z_j Z_k - B_{jk}$.

For convenience, let $D$ denote $D(\lambda)$ and $\bar D$ denote $D(\bar \lambda)$,
The result is
\begin{eqnarray*}
\hat L &=& Z^T \bar D W \bar D Z + \trace(D W D B) - \trace(\bar D W \bar D B).
\end{eqnarray*}

It follows that,
\begin{eqnarray*}
 L - \hat L 
  &=& \beta^T W \beta - 2 Z^T D W \beta + Z^T D W D Z -\\
   && \quad Z^T (I - D) W (I - D) Z - \trace((W - D W - W D) B)\\
  &=& \beta^T W \beta - 2 Z^T D W \beta + Z^T(D W + W D - W)Z  + \trace((D W + W D - W) B).
\end{eqnarray*}

Let $A = D W + W D - W$ and $C = 2 D W \beta$.
Then,
\begin{eqnarray*}
 \Var(L - \hat L)
  &=& \Var( Z^T A Z - Z^T C ) \\
  &=& \Var( Z^T A Z) + \Var(Z^T C) - 2 \Cov( Z^T A Z, Z^T C ) \\
  &=& 2\trace(A B A B) + \beta^T Q \beta,
\end{eqnarray*}
where
\begin{eqnarray*}
Q/4
 &=& A B A + W D B D W - 2 A B D W \\
 &=& (D W + W D - W) B (W D + D W - W) + W D B D W - 2 (D W + W D - W) B D W.
\end{eqnarray*}
Hence, plugging in unbiased estimates of the linear and quadratic forms
involving $\beta$, we get an estimate of the variance:
\begin{equation}
\hat\tau^2 = 2\trace(A B A B) + Z^T Q Z - \trace(Q B).
\end{equation}

\section*{Acknowledgments}

The authors would like to thank the Executive Editor for his careful reading
and helpful suggestions.

\section*{References}

Papers below with references to {\tt astro-ph} numbers can be found on-line in multiple formats
at {\tt http://xxx.lanl.gov/archive/astro-ph}.
In addition, the following websites have useful introductory material about WMAP and the
CMB:
\begin{enumerate}
\item The WMAP home page \par{\tt http://map.gsfc.nasa.gov/},
\item Wayne Hu's CMB Physics Pages
\par{\tt http://background.uchicago.edu/$\sim$whu/physics/physics.html}, 
\item Ned Wright's Cosmology Tutorial \par{\tt http://www.astro.ucla.edu/$\sim$wright/cosmolog.htm},
\item and an introductory review of CMB studies \par{\tt http://astron.berkeley.edu/$\sim$mwhite/rosetta/index.html}.
\end{enumerate}

\parskip=2ex

\noindent
{\sc Balbi, A., Baccigalupi, C., Perrotta, F., Matarrese, S., \& Vittorio, N.} (2003).
Probing Dark Energy with the Cosmic Microwave Background: Projected Constraints from the Wilkinson Microwave Anisotropy Probe and Planck
{\em Astrophys. J.}, {\bf 588}, L5-L8.

\noindent
{\sc Bennett et al. (2003).}
First Year Wilkinson Microwave Anisotropy Probe (WMAP) Observations:
Preliminary Maps and Basic Results
{\em Astrophys. J. Supp.} {\bf 148}, 1-27.

\noindent
{\sc Beran, R. (2000).}
REACT scatterplot smoothers:
superefficiency through basis economy.
{\em J. Amer. Statist. Assoc.} {\bf 95}, 155-171.

\noindent
{\sc Beran, R. and D\"umbgen, L.} (1998).
Modulation of estimators and confidence sets.
{\em Ann. Statist.} {\bf 26}, 1826-1856.

\noindent
{\sc Croft et al.} (2002). 
Toward a Precise Measurement of Matter Clustering: Ly-alpha Forest Data at Redshifts 2-4,
{\em Astrophys. J.}, {\bf 581}, 20--52

{\sc Gnedin, N. Y. and Hamilton, A. J. S.}, (2002).
Matter power spectrum from the Lyman-alpha forest: myth or reality?
\emph{Monthly Notices of the Royal Astronomical Society}, {\bf 334}, 107--116. 

\noindent
{\sc Halverson, N.W and Leitch, E.M. and Pryke, C. and 
              Kovac, J. and Carlstrom, J.E. and Holzapfel, W.L. and 
              Dragovan, M. and Cartwright, J.K. and Mason, B.S. and 
              Padin, S. and Pearson, T.J. and Shepherd, M.C. and 
              Readhead, A.C.S.} (2002).
{D}{A}{S}{I} First Results: A Measurement of the 
Cosmic Microwave Background Angular Power Spectrum.
{\em Astrophy. J.l}, {\bf 568}, 38-45.

\noindent
{\sc Hinshaw et al.} (2003).
First-Year Wilkinson Microwave Anisotropy Probe (WMAP) Observations: The Angular Power Spectrum
{\em Astrophys. J. Supp.}, {\bf 148}, 135-159.

\noindent
{\sc Hu, W.} (1999). CMB Anisotropies: A Decadal Survey,
in \emph{Birth and Evolution of the Universe: RESCEU 1999}, {\tt astro-ph/0002520}.

\noindent
{\sc Hu, W.} (2001). Physics of CMB Anisotropies, Warner Prize Lecture, American Astronomical Society Meeting, 
San Diego, Jan. 2001, \hfil\break{\small\tt http://background.uchicago.edu/$\sim$whu/Presentations/warnerprint.pdf}.

\noindent
{\sc Hu, W.} (2003). CMB Temperature and Polarization Anisotropy Fundamentals, 
\emph{Annals of Physics}, {\bf 303}, 203, {\tt astro-ph/0210696}.

\noindent
{\sc Hu, W. and Dodelson, S.} (2002). Cosmic Microwave Background Anisotropies, 
\emph{Ann. Rev. Astron. Astrophys.}, {\bf 40}, 171, {\tt astro-ph/0110414}.

\noindent
{\sc Hu, W. and Sugiyama, N.} (1995).
Toward Understanding CMB Anisotropies and Their Implications
{\em Phys. Rev. D}, {\bf 51}, 2199-2630.

\noindent
Knox, L. (1999). Cosmic Microwave Background Bandpower Windows Revisited, 
\emph{Physic Rev. D}, 60, issue 10, 103516, 5 pages.

\noindent
Kuo et al. (2003). 
High Resolution Observations of the CMB Power Spectrum with ACBAR,
{\tt astro-ph/0212289}.

\noindent
{\sc Lee, A. T. et. al.} (2002),
A High Spatial Resolution Analysis of the MAXIMA-1 Cosmic
Microwave Background Anisotropy Data,
{\em Astrophys. J.}, 561, L1-L6.

\noindent
Marinucci, D. (2004).
Testing For Non-Gaussianity On Cosmic Microwave Background Radiation: A Review
\emph{Statistical Science}, this issue.

\noindent
{\sc Mason et al.} (2002). 
The Anisotropy of the Microwave Background to l = 3500: Deep Field Observations with the Cosmic Background Imager, \emph{Astrophys. J.}, {\bf 591}, 540--555.

\noindent
{\sc Netterfield, C. B. et. al.} (2002).
A measurement by BOOMERANG of multiple peaks in the angular
power spectrum of the cosmic microwave background.
{\em Astrophys. J.}, {\bf 571}, 604-614.

\noindent
Pearson et al. (2003).
The Anisotropy of the Microwave Background to l = 3500: Mosaic Observations with the Cosmic Background Imager,
{\em Astrophys. J.}, {\bf 591}, 556--574.

\noindent
Percival et al. (2001).
The 2dF Galaxy Redshift Survey: the power spectrum and the matter content of the Universe,
\emph{Monthly Notices of the Royal Astronomical Society}, {\bf 327}, 1297--1306.

\noindent
Ruppert, D., Wand, M. and Carroll, R. (2003).
\emph{Semiparametric Regression.}
Cambridge University press. Cambridge.

\noindent
{\sc Schwarz, D.J.} (2003). The first second of the Universe, 
\emph{Annalen Phys.}, {\bf 12}, 220--270, {\tt astro-ph/0303574}.

\noindent
{\sc Seljak, U. and Zaldarriaga, M.} (1996).
A Line-of-Sight Integration Approach to Cosmic Microwave Background Anisotropies
{\em Astrophys. J.}, {\bf 469}, 437-444.

\noindent
{\sc Sievers et al.} (2003).
Cosmological Parameters from Cosmic Background Imager Observations and Comparisons with BOOMERANG, DASI, and MAXIMA,
\emph{Astrophys. J.}, {\bf 591}, 599--622. 

\noindent
{\sc Smoot, G. et al.} (1992).
Structure in the COBE differential microwave radiometer first-year maps.
{\em  Astrophys. J.}, {\bf 396L}, L1-L5.

\noindent
{\sc Spergel, D. et al.} (2003).
First-Year Wilkinson Microwave Anisotropy Probe (WMAP) Observations: Determination of Cosmological Parameters
{\em Astrophys. J. Supp.}, {\bf 148}, 175-194.
\noindent

\noindent
{\sc Stein, C.} (1981).
Estimation of the mean of a multivariate normal
distribution.
{\em Ann. Statist.} {\bf 9}, 1135-1151.

\noindent
{\sc Sun, J.} and {\sc Loader, C. R.} (1994).
Simultaneous Confidence Bands in Linear Regression and Smoothing,
\emph{Annals of Statistics}, {\bf 22}, 1328--1345.

\end{document}